\documentclass{article}
\usepackage{amssymb}
\usepackage{amsmath}
\usepackage{amsfonts}
\usepackage{amsthm}
\usepackage[T1]{fontenc}
\usepackage{authblk}

\newtheorem{definition}{Definition}
\newtheorem{proposition}{Proposition}

\newtheorem{theorem}{Theorem}

\newtheorem{remark}{Remark}

\begin{document}
\title{Toric data, Killing forms and complete integrability of geodesics in
Sasaki-Einstein spaces $Y^{p,q}$}

\author[1]{Vladimir Slesar\thanks{slesar.vladimir@ucv.ro}}
\author[2]{Mihai Visinescu\thanks{mvisin@theory.nipne.ro}}
\author[3,4]{Gabriel Eduard
V\^ilcu\thanks{gvilcu@upg-ploiesti.ro}\thanks{gvilcu@gta.math.unibuc.ro}}
\affil[1]{Department of Mathematics, University of Craiova,

Str. Al.I. Cuza, Nr. 13, Craiova 200585, Romania}
\affil[2]{Department of Theoretical Physics,

National Institute for Physics and Nuclear Engineering,

Magurele, P.O.Box M.G.-6, Romania}
\affil[3]{Department of Cybernetics and Economic Informatics,

Petroleum-Gas University of Ploie\c sti,

Bd. Bucure\c sti, Nr. 39, Ploie\c sti 100680, Romania}
\affil[4]{Faculty of Mathematics and Computer Science,

Research Center in Geometry, Topology and Algebra,

University of Bucharest, Str. Academiei, Nr. 14, Sector 1,
Bucharest 060042, Romania}
\date{\today }
\maketitle

\begin{abstract}

In the present paper we show that the
complete
list of special Killing forms on the $5$-dimensional Sasaki-Einstein spaces
$Y^{p,q}$ can be extracted using the
symplectic potential and the classical Delzant construction. The results
achieved here
agree with previous ones obtained by direct computation, proving the
reliability of
the method which stands in fact as a general algorithm for toric
Sasaki-Einstein manifolds.  Finally, we discuss the integrability of geodesic
motion in $Y^{p,q}$ spaces.

\end{abstract}

\section{Introduction}

Symmetries of dynamical equations have always played very important role in
physics.
The most common meaning of the symmetry is associated to that of isometry that
leaves the
metric invariant. A one-parameter continuous isometry is related to Killing
vectors.
More general one can consider a physical system evolving in a given background
and analyze the
symmetries of the whole phase space of the system such that the dynamics is
left invariant.
Such symmetries are often referred to as hidden symmetries whose conserved
quantities are
of higher order in momenta.

The conserved quantities polynomial in momenta are constructed making use of
St\"ackel-Killing
tensors which are natural generalization of the Killing vectors. There is also
an
antisymmetric generalization of the Killing vectors represented by Killing-Yano
tensors.
They play an important role in the study of complete integrability of geodesic
equations
and complete separation of variables for the Hamilton-Jacobi, Klein-Gordon and
Dirac
equations  in general Kerr-NUT-(A)dS metrics.

In the last time Sasaki-Einstein geometry has become of significant interest in
some modern
developments in mathematics and theoretical physics \cite{B-G-2008,Sp}. The
Sasaki-Einstein
spaces provide supersymmetric backgrounds relevant to the AdS/CFT
correspondence \cite{JM}.

In this paper we want to take a closer look at the special Killing forms on the
$5$-dimensional
toric Sasaki-Einstein $Y^{p,q}$ spaces.
The special Killing forms on the Sasaki-Einstein manifold $Y^{p,q}$ were
firstly obtained
by a direct calculation in \cite{Vis}. Later, in \cite{S-V-V},
the authors show that working with foliated coordinates we can also extract the
special Killing forms
on a Sasaki-Einstein manifolds, developing a general alternative approach which
was exemplified in the
case of the five-dimensional $Y^{p,q}$ spaces, obtaining the same results as in
\cite{Vis}.
Now we give a third approach, proving that the description of the Calabi-Yau
cone
$C(Y^{p,q})$ using toric data allows us to extract the special Killing forms on
$Y^{p,q}$ based
on a standard method.
Using the complete set of special Killing forms we construct the
St\"ackel-Killing tensors and the
corresponding conserved quantities, quadratic in momenta.
Finally we investigate the integrability of the geodesic motion and show that
the system is
completely integrable.

The paper has the following organization: In Section 2.1 we review some
well-known properties of
Killing tensors. In Section 2.2 we describe the special Killing forms on
Sasaki-Einstein manifolds. In Sections 2.3  and 2.4 we present the evaluation
of
symplectic potential and the holomorphic volume form in complex coordinates
using toric data.
In Section 3 we exemplify the general scheme in the case of the five
dimensional Sasaki-Einstein
spaces $Y^{p,q}$ and write down the symplectic and complex coordinates.
In Section 4 we present the complete set of special Killing forms on $Y^{p,q}$
spaces.
In Section 5 we evaluate the St\"ackel-Killing tensors constructed from
Killing-Yano tensors and
prove the complete integrability of geodesic motion in $Y^{p,q}$ spaces. The
paper ends with
conclusions in Section 6.

\section{Preliminaries}

\subsection{Killing forms and St\"ackel-Killing tensors}
The \emph{Killing forms} (sometimes called \emph{Killing-Yano tensors}) stand
as a
natural extension of classical Killing 1-forms (which are dual to Killing
vector fields).
We introduce these differentiable forms in accordance with \cite{Semm}.
Throughout the paper we use standard conventions: $\nabla$ is the Levi-Civita
connection with respect to the metric $g$, $X^*$ is the $1$-form dual
to the vector field $X$, $\lrcorner$ is the operator dual to the wedge
product.

\begin{definition}
If $(M,g)$ is a Riemannian manifold, then a Killing form of rank $p$
is a $p$-form $\psi$ which have to satisfy the equation
\begin{equation}\label{CKY}
\nabla_X\psi=\frac{1}{p+1}X \lrcorner d\psi \,,
\end{equation}
for any vector field $X$ on $M$.
\end{definition}

The corresponding equation of \eqref{CKY} in component notation becomes
\[
\nabla_{(j}\psi_{i_1)i_2 \dots i_p} = 0 \,.
\]
Here the round brackets are used to denote symmetrization over the indices
within.

It turns out that the most part of known interesting Killing forms satisfy for
some
constant $c$ the additional equation
\begin{equation} \label{def_special kill}
\nabla_X(d\psi) = c X^* \wedge \psi \,,
\end{equation}
for any vector field $X$ on $M$.
\begin{definition}
The particular class of tensors which satisfy the above relation are called
\emph{special Killing forms} {\rm \cite{Semm}}.
\end{definition}

We introduce below a symmetric generalization of the Killing vectors; if a
symmetric tensor
$K_{i_1 \cdots i_r}$ of rank $r>1$ satisfies the generalized Killing equation
\[
\nabla_{(j}K_{i_1 \cdots i_r)} =0\,,
\]
then it is called a \emph{St\"{a}ckel-Killing tensor}.

We remark that for any geodesic $\gamma$ with tangent vector $\dot{\gamma}^i$
\begin{equation}\label{SKcons}
Q_K =K_{i_1 \cdots i_r} \dot{\gamma}^{i_1} \cdots \dot{\gamma}^{i_r}\,,
\end{equation}
is constant along $\gamma$.

This stands as an analogue of the conserved quantities associated with Killing
vectors.
Given two Killing-Yano tensors $\psi^{i_1, \dots, i_k}$ and
$\sigma^{i_1, \dots, i_k}$ there is a St\"{a}ckel-Killing tensor of
rank $2$:
\begin{equation}\label{KYY}
K^{(\psi,\sigma)}_{ij} = \psi_{i i_2 \dots i_k}
\sigma_{j}^{\phantom{j}i_2 \dots i_k}+ \sigma_{i i_2 \dots i_k}
\psi_{j}^{\phantom{j}i_2 \dots i_k} \,.
\end{equation}

This result represents an important connection between these
two generalizations of the Killing vectors, and offers a method to generate
higher
order integrals of motion by identifying the complete set of Killing forms.

\subsection{Special Killing forms on Sasaki-Einstein manifolds}
We pass now to a remarkable class of manifolds where special Killing forms are
known to exist;
these are Sasaki-Einstein manifolds.
In the following we give a brief presentation. First we introduce the metric
cone $C(M)$ of
the manifold $M$. This is in fact the product manifold $M\times
\mathbb{R}_{>0}$,
with $\dim C(M)=2n+2$, endowed with the warped metric $g_{cone}:=dr^2+r^2g$.

\begin{definition}
A differentiable manifold $M$ is called \emph{Sasaki manifold} if its metric
cone $C(M)$
is a K\"ahler manifold.
\end{definition}

If $\mathcal{J}$ represents the complex structure on the cone manifold, then we
denote
\[
\tilde{K}:=\mathcal{J}{(r \frac{\partial}{\partial r})}.
\]

Considering the restriction of $\tilde{K}$ to the submanifold determined by the
condition
$r=1$ we obtain the Reeb vector field $\mathcal{B}$
on the Sasaki manifold $M$. The dual 1-form of $\mathcal{B}$ on $M$ is denoted
by $\eta$,
and extending on the cone manifold $C(M)$ we obtain that the dual of
$\tilde{K}$ is $r^2\eta$.
Now the K\"ahler form $\omega$ can be expressed as
\[
\omega=\frac{1}{2} d(r^2\eta)=\frac{1}{2}\mathrm{i}\partial\bar{\partial} r^2.
\]
where $\mathbf{i}:=\sqrt{-1}$ and $\partial$ and $\bar{\partial}$ are the
canonical
operators associated to $\mathcal{J}$.

From here we easily see that $F:=\frac{r^2}{4}$ is the K\"ahler potential.

\begin{definition}
An \emph{Einstein manifold} is a Riemannian manifold $(M,g)$ satisfying the
Einstein condition
\begin{equation}\label{Einstein}
Ric_{g} = \lambda g \,,
\end{equation}
for a real constant $\lambda$, where $Ric_{g}$ denotes the Ricci
tensor of $g$.
\end{definition}

Einstein manifolds with $\lambda=0$ are called
\emph{Ricci-flat manifolds}. Finally, a \emph{Sasaki-Einstein manifold} is a
Riemannian manifold $(M,g)$ that is both Sasaki and Einstein. We note that
in the case of Sasaki-Einstein manifolds one always has (\ref{Einstein})
with the Einstein constant $\lambda=2n$. It turns out that a Sasaki manifold
$M$ is
Einstein if and only if the metric cone $C(M)$ is  K\"{a}hler Ricci-flat.
Consequently, the metric cone of a Sasaki-Einstein manifolds will be a K\"ahler
manifold with
flat Ricci tensor, i.e. a Calabi-Yau manifold (see e.g. \cite{B-G-2008}).

\begin{definition}
A \emph{toric Sasaki manifold} $M$ is a Sasaki manifold
whose K\"{a}hler cone $C(M)$ is a toric K\"{a}hler manifold {\rm \cite{Gu}}.
\end{definition}

For instance, a five-dimensional toric Sasaki-Einstein manifold is a
Sasaki-Einstein manifold with three $U(1)$ isometries. Then, using symplectic
geometry of the cone $C(M)$ one can introduce canonical coordinates and the
Sasaki-Einstein structure can be described in terms of toric data together
with a single function $G$, a symplectic potential (see e.g. \cite{O-Y}).
Remarkable examples are represented by the spaces $Y^{p,q}$ with relative
prime numbers $p$ and $q$, which topologically are $S^1$\--fibration over
$S^2 \times S^2$.

Next, we discuss the existence of special Killing forms on Sasaki-Einstein
manifolds. In this respect, a remarkable
correspondence between special Killing forms defined on such manifold $M$
and the parallel forms
defined on the corresponding metric cone $C(M)$ was stated by Semmelman
\cite{Semm}.

\begin{theorem}{\rm\cite{Semm}}
A $p-$dimensional differential form $\Psi $ is a special Killing form on $M$ if
and only if the corresponding form
\begin{equation}
\Psi _{cone}:=r^pdr\wedge \Psi +\frac{r^{p+1}}{p+1}d\Psi \,,
\label{eq Semmelmann}
\end{equation}
is parallel on $C(M)$.
\end{theorem}

Using the defining equation \eqref{def_special kill}  we can show that on a
five-dimensional Sasaki manifold $M$ two special Killing forms can be directly
written employing the
$1-$form $\eta$ \cite{Semm}:
\begin{equation}\label{sKPsi}
\Psi_1 = \eta \wedge d\eta,\quad  \Psi_2=\eta \wedge (d\eta)^2 \,.
\end{equation}
Other two closed conformal Killing forms called $\ast$-Killing forms, are also
obtained:
\begin{equation}\label{sKPhi}
\Phi_1 = d\eta,\quad \Phi_2=(d\eta)^2 \,.
\end{equation}
In the case of the Calabi-Yau cone $C(M)$, with the method offered in
\cite{Semm}
two additional special Killing forms on $M$ can be extracted using parallel
forms of the cone. We introduce now the \emph{holomorphic complex volume form}.

\begin{definition}
If $\mathrm{Vol}$ is the volume form on the metric cone, then the holomorphic
volume
form $\Omega$ is defined by the relation
\[
\mathrm{Vol}=\frac{1}{(n+1)!} \omega ^{n+1}=\frac{i^{n+1}}{2^{n+1}}
(-1)^{n(n+1)/2}\Omega \wedge \bar{\Omega} \,.
\]
\end{definition}
We recall now the following classical fact.
\begin{theorem}{\rm\cite[Chapter 17]{Mor}}
On Ricci flat K\"ahler manifolds the holomorphic complex form is parallel.
\end{theorem}

Consequently, in our setting the holomorphic complex volume form $\Omega$ of
$C(M)$ and its conjugate \cite{Semm} are the
two needed parallel forms.

\subsection{Symplectic potential and complex coordinates}
In order to write the holomorphic complex form we need complex coordinates and
corresponding metric coefficients on the cone manifold.
In the previous papers \cite{Vis,S-V-V} these coordinates were obtained by a
direct computation. In order to show that
this can be done using the classical Delzant construction and symplectic
potential
we start out by considering first the symplectic (action-angle) coordinates
$(y ^i,\Phi ^i)$. The angular coordinates $\Phi ^i$ will
generate the toric action. In order to obtain the $y^i$ the key ingredient is
represented by the momentum map $\mu =\frac 12r^2\eta $. We have the
correspondence
\begin{equation}
y^i=\mu (\partial /\partial \Phi ^i)\,. 
\end{equation}
The corresponding K\"ahler metric on the cone $C(M)$ is constructed
using the \emph{symplectic potential} $G$ \cite{M-S-Y, Abr}.

For this purpose we briefly present below the Delzant result.

A \emph{Delzant polytope} is a convex polytope such that there are $n$ edges
meeting at each vertex, each edge meeting at the vertex is of form $1+tu_i$,
where $u_i\in \mathbb{Z}^n$, and $\{u_i\}$ can be chosen to form a basis in
$\mathbb{Z}^n$. This polytope can be described by the inequalities
\[
l_A(y):=\left\langle y,v_A\right\rangle \ge 0\mbox{,\, for }1\le A\le d\,,
\]
where $\{v_A\}$ are inward pointing normal vectors to the facets of the
polytope and $d$ is the number of facets \cite{Abr, Gu}.

It is possible to associate to any Delzant polytope $P\in \mathbb{R}^n$
a close connected symplectic manifold $M$, together with a Hamiltonian
$\mathbb{T}^n$
action on the manifold. In fact it can be shown that the polytope turns out to
be the image
of the momentum map, $P=\mu (M)$ (see e.g. \cite{Ler}).

\begin{remark}
In the case of the Calabi-Yau cone we take $C(M)$ to be {\it Gorenstein}
which is a necessary condition to admit a Ricci-flat K\"{a}hler metric and $M$
to admit a Sasaki-Einstein metric.
For affine toric varieties it is well-known that $C(M)$ being Gorenstein is
equivalent
to the existence of a basis for the torus $\mathbb{T}^n$ for which
\begin{equation}\label{Gorenstein}
v_a=(1,w_a)\,,
\end{equation}
for each $a=1,\cdots,d$ and $w_a \in\mathbb{Z}^{n-1}$ {\rm \cite{M-S,M-S-Y}}.
\end{remark}

The symplectic potential $G$ can be written in terms of the toric
data \cite{M-S-Y}
\begin{equation}
G=G^{can}+G^{\mathcal{B}}+h.  \label{G}
\end{equation}
In the above description of $G$ the first terms is the canonical symplectic
potential and it is computed
using the corresponding Delzant polytope.
\[
G^{can} =\frac 12\sum_Al_A(y )\log l_A(y),  \label{G_can_G_B} \\
\]
The second term is related to the Reeb vector field $\mathcal{B}$. Here
we define the affine function $l_{\mathcal{B}}:=\left\langle {\mathcal{B}},\cdot
\right\rangle $, and $l_\infty :=\left\langle \sum_Av_A,\cdot \right\rangle $
and $G^{{\mathcal{B}}}$ is written as
\[
G^{\mathcal{B}} =\frac 12l_{\mathcal{B}}(y)\log l_{\mathcal{B}}(y)-\frac
12l_\infty (y)\log l_\infty (y)\,.
\]

Finally, as in the general case $G$ needs to satisfy the Monge-Amp\`ere
equation,
a homogeneous function $h$ of degree $1$ in variables $y$ is added.

From equation \eqref{G} it becomes clear the relevance of the Reeb vector field
$\mathcal{B}$ and the function $h$.
Note that ${\mathcal{B}}$ is constant \cite{M-S-Y}. We make here some
references
regarding the possibility to calculate this vector.
According to the AdS/CFT correspondence the volume of the Sasaki-Einstein space
corresponds to the central charge of the dual conformal field theory.
There are two known different algebraic procedures to extract the components of
the Reeb vector from the toric data. The first procedure is based on \emph{the
maximization of the central charge} ($a$-maximization) \cite{Int-Wec} used in
connection with the computation of the Weyl anomaly in 4-dimensional field
theory. The
second one is known as \emph{volume minimization} (or $Z$-minimization)
\cite{M-S-Y}.

Using now the symplectic potential we can write the metric on the K\"ahler
manifold with
respect to the symplectic coordinates
\[
ds^2=G_{ij}dy^idy^j+G^{ij}d\Phi ^id\Phi ^j\,,
\]
where the metric coefficients are computed
\[
G_{ij}=\frac{\partial ^2G}{\partial y^i\partial y^j}\,,
\]
with $\left( G^{ij}\right) :=\left( G_{ij}\right) ^{-1}$.

The main outcome of the above construction is that it allows us to pass to
the complex coordinates $z^i:=x^i+\mathrm{i}\Phi ^i$.

Considering the K\"ahler potential $F$ as the Legendre dual of $G$
\[
F(x)=\left( y^i\frac{\partial G}{\partial y^i}-G\right) :
\left(y=\partial F/\partial x\right) \,.
\]
we can obtain the coordinates $x^i$ using the Legendre transform
\[
x^i=\frac{\partial G}{\partial y^i},\mbox{\,\,}y^i=\frac{\partial F}{\partial
x^i}\,.
\]

The metric structure is now written with respect to the coordinate patch $(x^i,\Phi^i)$
in the following manner
\[
ds^2=F_{ij}dx^idx^j+F_{ij}d\Phi ^id\Phi ^j\,,
\]
where the metric coefficients are again obtained using the Hessian of the
K\"ahler potential $F$, i.e.
\[
F_{ij}=\frac{\partial ^2F}{\partial x^i\partial x^j}\,.
\]
Another useful remark is \cite{Abr}
\begin{equation}\label{G-F}
\left( F_{ij}\right) =\left( G^{ij}\right) \,.
\end{equation}

Using (\ref{G}), we are able to express the coordinates $x^i$ and the
metric coefficients $G_{ij}$
\begin{equation*}
\begin{split}
x^i &=\frac{\partial G}{\partial y^i}=\frac 12\sum_Av_A^i\log l_A(y)
+\frac 12{\mathcal{B}}^i(1+\log l_{\mathcal{B}}(y)) \\
&-\frac 12\sum_Av_A^i\log l_\infty (y)+\lambda _i   \,,\\
G_{ij} &=\frac 12\sum_A\frac{v_A^iv_A^j}{l_A(y)}
+\frac 12\frac{{\mathcal{B}}_i{\mathcal{B}}_j}{l_{\mathcal{B}}(y)}
-\frac 12\frac{\sum_Av_A^i\sum_Av_A^j}{l_\infty (y)}\,.
\end{split}
\end{equation*}

\subsection{The holomorphic volume form in complex coordinates}

As we can calculate the complex coordinates and the metric coefficients in
the above manner, we can use them to express
the holomorphic volume form.

Eventually ignoring the multiplicative constant, $\Omega $ can be written as
\cite{M-S-Y}
\[
\Omega =\exp({i\alpha })\det (F_{ij})^{1/2}dz^1\wedge ..\wedge dz^n \,.
\]

We use in the following the fact that the Calabi-Yau metric cone is Ricci
flat. From the classical formula
\[
\rho =-\mathrm{i}\partial \bar \partial \log \det (F_{ij})\,,
\]
using \eqref{G-F}, by a simple computation we get (see \cite{M-S-Y})
\begin{equation}
\det (G_{ij})=\exp{\left(2\gamma _i\frac{\partial G}{\partial y^i}-c\right)}\,,
\label{det_G}
\end{equation}
with constants $\gamma _i$, and $c$.

Now, as the metric has to be smooth, from (\ref{det_G}) it turns out that
\cite{O-Y}
\[
\gamma =(-1,0,..,0) \,,
\]
and, consequently, the Hessian of the K\"ahler potential is written
\[
\det (F_{ij})=\exp{\left(2x^1+c\right)}.
\]

Plugging this final result in the above formula, we get
\[
\Omega =\exp({x^1+i\alpha})dz^1\wedge ..\wedge dz^n\,.
\]

Finally, $\Omega $ should be closed (as it is parallel), and we
can choose the phase $\alpha $ to be $\Phi ^1$. We obtain the desired result.
\begin{proposition} {\rm \cite{M-S-Y}}
With respect to the complex coordinates $(z^i)$ the holomorphic volume form
has the following simple form
\begin{equation}
\Omega =\exp (z^1)dz^1\wedge ..\wedge dz^n.  \label{Omega}
\end{equation}
\end{proposition}
In the next sections we use this relation in order to extract the special
Killing forms on manifolds of Sasaki-Einstein type.

\section{Symplectic and complex coordinates on $Y^{p,q}$}

The metric tensor of the 5-dimensional
$Y^{p,q}$ manifold can be written as \cite{M-S}
\begin{equation}
\begin{split}\label{Ypq}
ds^2 & = \frac{1-c\, y}{6}( d \theta^2 + \sin^2 \theta\, d \phi^2)
+  \frac{1}{w(y)q(y)} dy^2
+ \frac{q(y)}{9} ( d\psi - \cos \theta \, d \phi)^2\\
& \quad  +
w(y)\left[ d\alpha + \frac{ac -2y+ c\, y^2}{6(a-y^2)}
[d\psi - \cos\theta \, d\phi]\right]^2\,,
\end{split}
\end{equation}
where
\begin{equation}
\begin{split}
w(y) & = \frac{2(a-y^2)}{1-cy}\,,\\
q(y) & = \frac{a-3y^2 + 2c y^3}{a-y^2}\,.
\end{split}
\end{equation}

This metric is Einstein with $Ric_g = 4 g$ for all values of the
constants $a,c$. Moreover the space is also Sasaki. For $c=0$ the
metric takes the local form of the standard homogeneous metric on
$T^{1,1}$ \cite{M-S}. Otherwise the constant $c$ can be rescaled by a
diffeomorphism and in what follows we take $c=1$. For
$
0 < \alpha < 1\,,
$
we can take the range of the angular coordinates $(\theta, \phi, \psi)$
to be $0\leq \theta \leq 2\pi\,, 0\leq \phi \leq 2\pi\,, 0\leq \psi \leq
2\pi$. Choosing $0 < a < 1$  the roots $y_i$  of the cubic equation
\begin{equation}\label{cubic}
a - 3 y^2 + 2 y^3 = 0 \,,
\end{equation}
are real, one negative $(y_1)$ and two positive $(y_2, y_3)$. If the
smallest of the positive roots is $y_2$, one can take the range of the
coordinate $y$ to be
\[
y_1\leq y \leq y_2 \,.
\]

We note that the parameter
$a$ of the cubic equation (\ref{cubic}) can be expressed in terms of the two
relatively prime positive integers
$p$ and $q$ as
\begin{equation}\label{R1}
a=\frac{1}{2}-\frac{p^2-3q^2}{4p^3}\sqrt{4p^2-3q^2},
\end{equation}
and the roots of the cubic $a - 3 y^2 + 2 y^3$ are
\begin{equation}\label{R2}
y_{1}=\frac{1}{4p}\left(2p-3q-\sqrt{4p^2-3q^2}\right),
\end{equation}
\begin{equation}\label{R3}
y_{2}=\frac{1}{4p}\left(2p+3q-\sqrt{4p^2-3q^2}\right),
\end{equation}
\begin{equation}\label{R4}
y_{3}=\frac{1}{2}+\frac{\sqrt{4p^2-3q^2}}{2p}.
\end{equation}

We can now define (see \cite{M-S}):
\begin{equation}\label{ag}
\alpha \equiv \ell \gamma\,,
\end{equation}
with
\begin{equation}\label{LL}
\ell = \frac{q}{ 3 q^2 - 2 p^2 + p\sqrt{4p^2 - 3q^2}}\,,
\end{equation}
and the following change of variables \cite{M-S}
\begin{equation}\label{sv}
\alpha = - \beta/6 - \psi^{\prime}/6 \,,\, \psi =  \psi^{\prime}\,.
\end{equation}

Therefore the
metric (\ref{Ypq}) takes the local Sasaki-Einstein form
\[
ds^2=ds^2(B_4)+w(y)[d\alpha+A]^2,
\]
where $B_4=S^2\times S^2$, $ds^2(B_4)$ is the non-trivial metric on $B_4$
described in \cite{GMSW} and $A$ is a 1-form given by
\begin{equation}
A = f(y)(d\psi - \cos\theta d\phi)\,,
\end{equation}
where
\begin{equation}
f(y) =  \frac{a -2y+  y^2}{6(a-y^2)}\,.
\end{equation}

Moreover, the Reeb vector ${\mathcal{B}}$ and the dual
1-form $\eta$ are given by \cite{M-S}:
\begin{equation}\label{Reeb}
{\mathcal{B}} = 3 \frac {\partial}{\partial \psi^{\prime}} = 3 \frac
{\partial}{\partial \psi}
- \frac{1}{2\ell}\frac {\partial}{\partial \gamma}=
3 \frac {\partial}{\partial \psi}
- \frac{1}{2}\frac {\partial}{\partial \alpha}\,
\end{equation}
and
\begin{equation}
\eta = -2y (d\alpha +A) + \frac13 q(y) (d\psi - \cos\theta d\phi)\,.
\end{equation}

We remark now that the 1-form $\eta$ can be written in a simple way as:
\begin{equation}
\begin{split}\label{eta}
\eta &= -2y d\alpha + \frac{1-y}3 (d\psi - \cos\theta d\phi)\\
&= -2y \ell d\gamma + \frac{1-y}3 (d\psi - \cos\theta d\phi)\,,
\end{split}
\end{equation}
and it is easy to see that $\eta(\mathcal{B}) = 1$.

In relation to the angular variables $\phi,\psi,\gamma$, the basis for an
effectively acting
$\mathbb{T}^3$ action is \cite{M-S}
\begin{equation}\label{be}
\begin{split}
e_1& =\frac{\partial}{\partial \phi}+\frac{\partial}{\partial \psi}\,, \\
e_2& =\frac{\partial}{\partial \phi} -\frac{l}{2}\frac{\partial} {\partial
\gamma}\,, \\
e_3& =\frac{\partial}{\partial \gamma }\,,
\end{split}
\end{equation}
with $l= p-q$.

It is easy to see now that, considering this basis, the  Reeb vector has the
components
\begin{equation}
{\mathcal{B}} = \left(3, -3, -\frac32 (l + \frac1{3\ell})\right)\,.
\end{equation}

If we write now the basis \eqref{be} in the following form
\begin{equation}\label{ePhi}
e_i=\frac \partial {\partial \Phi ^i}\,,
\end{equation}
then we obtain after some standard computations that
\begin{equation}
\left(
\begin{array}{c}
\frac \partial {\partial \phi}\\
\frac \partial {\partial \psi}\\
\frac \partial {\partial \gamma}
\end{array}
\right)=
\left(
\begin{array}{ccc}
0 & 1 & \frac l2 \\
1 & -1 & - \frac l2 \\
0 & 0 & 1
\end{array}
\right)
\left(
\begin{array}{c}
\frac \partial {\partial \Phi^1}\\
\frac \partial {\partial \Phi^2}\\
\frac \partial {\partial \Phi^3}
\end{array}
\right)
\,,
\end{equation}
and therefore we derive
\begin{equation}\label{Phi}
\begin{split}
\Phi^1& = \psi\,, \\
\Phi^2& = \phi -\psi\,, \\
\Phi^3& =\frac l2 \phi - \frac l2 \psi + \gamma\,.
\end{split}
\end{equation}

Using now the correspondence between the symplectic (action-angle) coordinates
$(y^i,\Phi ^i)$ and the momentum map $\mu$:
\begin{equation}
y^i=\mu (\partial /\partial \Phi ^i) =
\frac{r^2}2 \eta(\partial /\partial \Phi ^i)\,, \label{yi}
\end{equation}
we deduce that, in this new basis \eqref{ePhi}, the momentum map becomes
\begin{equation}\label{mm}
\vec{y} = (y^1,y^2,y^3) = \left[\frac{r^2}6 (1-y)(1 -\cos\theta)\,,\,
-\frac{r^2}6 (1-y)\cos\theta + \frac{r^2}2 l\ell y\,,\, -\ell r^2 y\right]\,.
\end{equation}

Next, in order to introduce the complex coordinates on the Calabi-Yau cone
$C(Y^{p,q})$, we need the symplectic potential $G$. First of all, let us
consider
the toric data for $Y^{p,q}$  \cite{M-S,O-Y}):
\begin{equation}\label{v14}
v_1=[1,-1,-p]\,,\,v_2=[1,0,0]\,,\,v_3=[1,-1,0]\,,\,v_4=[1,-2,-p+q]\,.
\end{equation}

Now, from \eqref{G} we remark that the symplectic potential in the case of
$Y^{p,q}$
contains the function $h$, in contradistinction to the case of the homogeneous
Sasaki-Einstein manifold $T^{1,1}$ (see \cite{S-V-V1,S-V-V2}). However, it can
be
proved that one can derive a more simple expression \cite{O-Y}
\begin{equation}
G=\sum_{A=1}^6\frac 12\left\langle v_A,y \right\rangle \log\left\langle v_A,y
\right\rangle \,,
\end{equation}
by introducing two additional vectors $v_5$ and $v_6$ as follow:
\begin{equation}
\begin{split}\label{v56}
v_5 := & {\mathcal{B}}- v_1- v_3 = \left(1, -1, - \frac12 p + \frac32 q -
\frac 1{2\ell}\right)\,,\\
v_6 := & -v_2 -v_4 = (-2, 2, p-q)\,.
\end{split}
\end{equation}

Now we note that the complex coordinates are
\begin{equation}
\begin{split}
z^1 =& x^1 + \mathrm{i}\psi\,,\\
z^2 =& x^2 + \mathrm{i}(\phi - \psi)\,,\\
z^3 =& x^3 + \mathrm{i}(\frac l2 \phi - \frac l2 \psi + \gamma)\,.
\end{split}
\end{equation}

In order to derive the expression of $x^i$ we use:
\begin{equation}\label{xi}
x^i=\frac{\partial G}{\partial y^i}= \frac 12\sum_1^6 v_{A}^{i}\log
\langle v_A,y \rangle +\frac 12\sum_1^6 v_{A}^{i}\,.
\end{equation}

In the sequel, for the sake of simplicity we will ignore the additive constants.
Using now \eqref{mm}, (\ref{v14}) and (\ref{v56})  in (\ref{xi}) and taking
account
of (\ref{R1}), (\ref{R2}), (\ref{R3}), (\ref{R4}) and  (\ref{LL}), we derive
after
some long but relatively straightforward algebraic calculations:
\begin{equation}
\begin{split}
x^1 =& 3\log
r+\log\sin\theta+\frac12\log\left(y^3-\frac{3}{2}y^2+\frac{a}{2}\right)\,,\\
x^2 =& -3\log
r-2\log\cos\frac{\theta}{2}-\frac12\log\left(y^3-\frac{3}{2}y^2+\frac{a}{2}
\right)\,,\\
x^3 =& \frac{p(y_1-y_3)}{1-y_1}\log
r-l\log\cos\frac{\theta}{2}\,\\&+\frac{p(1-y_3)}{2(1-y_1)}\log(y-y_3)-\frac
p2\log(y-y_1)\,,
\end{split}
\end{equation}
Hence we can introduce on $C(Y^{p,q})$ the following
patch of complex coordinates
\begin{equation}\label{Z}
\begin{split}
z^1 =& \log \left(r^3\sin\theta\sqrt{y^3-\frac{3}{2}y^2+\frac{a}{2}}\right) +
\mathrm{i}\psi\,,\\
z^2 =& -\log\left(
r^3\cos^2\frac{\theta}{2}\sqrt{y^3-\frac{3}{2}y^2+\frac{a}{2}}\right) +
\mathrm{i}(\phi - \psi)\,,\\
z^3 =&  \log
\frac{r^{\frac{p(y_1-y_3)}{1-y_1}}(y-y_3)^{\frac{p(1-y_3)}{2(1-y_1)}}}
{\left(\cos\frac{\theta}{2}\right)^l\sqrt{(y-y_1)^p}} + \mathrm{i}\left(\frac
l2 \phi -
\frac l2 \psi + \gamma\right)\,.
\end{split}
\end{equation}

\section{Special Killing forms on $Y^{p,q}$}

In this section we will prove that the patch of complex coordinates (\ref{Z})
obtained above are the perfect
ingredient in order to extract the special Killing forms on $Y^{p,q}$.

From \eqref{Z}, we obtain
\[
\begin{split}
& \exp( {z^1})=r^3\sin\theta\sqrt{y^3-\frac{3}{2}y^2+\frac{a}{2}}\exp
(\mathrm{i}{\psi })\,,
\\
& dz^1=\frac 3rdr+T_1, \\
& dz^2=-\frac 3rdr+T_2\,, \\
& dz^3=\frac{p(y_1-y_3)}{r(1-y_1)}dr+T_3\,.
\end{split}
\]
where
\begin{equation}\label{T}
\begin{split}
T_1&:=\cot \theta d\theta+\frac 12
\frac{3y^2-3y}{y^3-\frac{3}{2}y^2+\frac{a}{2}}dy+\mathrm{i}d\psi ,\\
T_2 &:=\tan \frac{\theta}{2} d\theta-\frac
12\frac{3y^2-2y}{y^3-\frac{3}{2}y^2+\frac{a}{2}}dy
+\mathrm{i}(d\phi - d\psi), \\ 
T_3 &:=\frac l2\tan \frac{\theta}{2}
d\theta+\frac{p(y_1-y_3)(y-1)}{2(1-y_1)(y-y_1)(y-y_3)}dy\\
&\ \ +\mathrm{i}\left(\frac l2 d\phi - \frac l2 d\psi + d\gamma\right)\,.
\end{split}
\end{equation}

Therefore, the holomorphic volume form is
\begin{equation}
\begin{split}
\Omega =&\exp (z^1)dz^1\wedge dz^2\wedge dz^3
=r^3\sin\theta\sqrt{y^3-\frac{3}{2}y^2+\frac{a}{2}}\exp (\mathrm{i}{\psi })\\
&\times\left(\frac 3rdr+T_1\right)\wedge \left(-\frac 3{r}dr+T_2\right)\wedge
\left(\frac{p(y_1-y_3)}{r(1-y_1)}dr+T_3\right) \,. \nonumber
\end{split}
\end{equation}

In our particular framework the equation (\ref{eq Semmelmann}) becomes
\[
\Omega =r^2dr\wedge \Psi +\frac{r^3}3d\Psi \,.
\]

In order to extract $\Psi$ we have to keep the trace of the differential
form $dr$ in the above equation. We derive

\begin{equation}
\Psi =\sin\theta\sqrt{y^3-\frac{3}{2}y^2+\frac{a}{2}}\exp (\mathrm{i}{\psi
})\left(3T_2\wedge
T_3+3T_1\wedge T_3+\frac{p(y_1-y_3)}{1-y_1}T_1\wedge T_2\right)\,.  \label{TT}
\end{equation}

Next, we will compute the wedge products from \eqref{TT} using \eqref{T}. After
some long but standard computations, we obtain
\begin{equation}
\begin{split}
T_2\wedge T_3
&=\tan\frac{\theta}{2}\left[\frac{p(y_1-y_3)(y-1)}{2(1-y_1)(y-y_1)(y-y_3)}+
\frac l4 \frac{3y(y-1)}{y^3-\frac{3}{2}y^2+\frac{a}{2}}\right]d\theta\wedge dy
\label{T2T3} \\
&\ -\mathrm{i}\left[\frac{p(y_1-y_3)(y-1)}{2(1-y_1)(y-y_1)(y-y_3)}+\frac l4
\frac{3y(y-1)}{y^3-
\frac{3}{2}y^2+\frac{a}{2}}\right]dy\wedge d\phi \\
&\ +\mathrm{i}\left[\frac{p(y_1-y_3)(y-1)}{2(1-y_1)(y-y_1)(y-y_3)}+\frac
l4\frac{3y(y-1)}{y^3-
\frac{3}{2}y^2+\frac{a}{2}}\right]dy\wedge d\psi \\
&\ -\frac{\mathrm{i}}{2}\frac{3y(y-1)}{y^3-\frac{3}{2}y^2+\frac{a}{2}}dy\wedge
d\gamma+
\mathrm{i}\tan\frac{\theta}{2}d\theta\wedge d\gamma\\
&\ -d\phi\wedge d\gamma+d\psi\wedge d\gamma\,,
\end{split}
\end{equation}
\begin{equation}
\begin{split}
T_1\wedge T_3 &=
\left[\frac{p(y_1-y_3)(y-1)}{2(1-y_1)(y-y_1)(y-y_3)}\cot\theta-
\frac l4\frac{3y(y-1)}{y^3-\frac{3}{2}y^2+\frac{a}{2}}\tan\frac
{\theta}{2}\right]d\theta\wedge dy\label{T1T3} \\
&  -\mathrm{i}\left[\frac{p(y_1-y_3)(y-1)}{2(1-y_1)(y-y_1)(y-y_3)}+\frac l4
\frac{3y(y-1)}{y^3-\frac{3}{2}y^2+\frac{a}{2}}\right]dy\wedge d\psi \\
&  +\mathrm{i}\frac l2 \cot\theta d\theta\wedge d\phi+\mathrm{i}\cot\theta
d\theta\wedge d\gamma-\mathrm{i}\frac l{2\sin\theta}d\theta\wedge d\psi\\
& +\mathrm{i}\frac l4\frac{3y(y-1)}{y^3-\frac{3}{2}y^2+\frac{a}{2}}dy\wedge
d\phi+\frac{\mathrm{i}}{2}\frac{3y(y-1)}{y^3-\frac{3}{2}y^2+\frac{a}{2}}
dy\wedge d\gamma\\
&-\frac l2 d\psi\wedge d\phi-d\psi\wedge d\gamma  \,,
\end{split}
\end{equation}
and
\begin{equation}
\begin{split}
\label{T1T2}
T_1\wedge T_2 &=-\frac {1}{2\sin\theta}\frac{3y(y-1)}{y^3-\frac{3}{2}y^2+
\frac{a}{2}}d\theta\wedge dy+\mathrm{i}\cot\theta d\theta\wedge d\phi\\
&\ -\mathrm{i}\frac{1}{\sin\theta}d\theta\wedge d\psi+\frac{\mathrm{i}}{2}
\frac{3y(y-1)}{y^3-\frac{3}{2}y^2+\frac{a}{2}}dy\wedge d\phi-d\psi\wedge
d\phi\,.
\end{split}
\end{equation}

Introducing now \eqref{T2T3}-\eqref{T1T2} in \eqref{TT} and ignoring the
multiplicative
constants, we derive
\begin{equation}\label{cvol}
\begin{split}
\Psi =&\sqrt{y^3-\frac{3}{2}y^2+\frac{a}{2}}\exp({\mathrm{i}\psi}) \\
&\times\left[ \left(a(y)d\theta\wedge dy
+\frac{1}{2\ell}\sin\theta d\psi\wedge d\phi-3\sin\theta d\phi\wedge
d\gamma\right) \right. \\
&\left.+\mathrm{i}\left(\frac{1}{2\ell}d\theta\wedge d\psi+3d\theta\wedge
d\gamma-a(y)\sin\theta dy\wedge d\phi-
\frac{1}{2\ell}\cos\theta d\theta\wedge d\phi\right) \right] \,,
\end{split}
\end{equation}
where
\[
\begin{split}
a(y)&=\frac{3p(y_1-y_3)(y-1)}{2(1-y_1)(y-y_1)(y-y_3)}-\frac{p(y_1-y_3)}{2(1-y_1)
}
\frac{3y(y-1)}{y^3-\frac{3}{2}y^2+\frac{a}{2}}\\
&=\frac{3py_2(y_1-y_3)}{1-y_1}\frac{1-y}{2y^3-3y^2+a}\\
&=-\frac{3}{2\ell}\frac{1-y}{2y^3-3y^2+a}.
\end{split}
\]

Now, we can easily obtain the real special Killing forms computing the real and
imaginary part of $\Psi $:
\[
\begin{split}
\Re \Psi  =
&\sqrt{y^3-\frac{3}{2}y^2+\frac{a}{2}}\left[\cos\psi\left(a(y)d\theta\wedge dy
+ \frac{1}{2\ell}\sin\theta d\psi\wedge d\phi-3\sin\theta d\phi\wedge
d\gamma\right)\right.\\
& \left. -\sin\psi\left(\frac{1}{2\ell} d\theta\wedge d\psi+3d\theta\wedge
d\gamma-a(y)\sin\theta dy\wedge d\phi-
\frac{1}{2\ell}\cos\theta d\theta\wedge d\phi\right)\right]\,,
\end{split}
\]
\[
\begin{split}
\Im \Psi  = &\sqrt{y^3-\frac{3}{2}y^2+\frac{a}{2}}\left[\sin\psi
\left(a(y)d\theta\wedge dy
+ \frac{1}{2\ell}\sin\theta  d\psi\wedge d\phi-3\sin\theta d\phi\wedge
d\gamma\right)\right.\\
& \left. +\cos\psi\left(\frac{1}{2\ell} d\theta\wedge d\psi+3d\theta\wedge
d\gamma-a(y)\sin\theta dy\wedge d\phi-
\frac{1}{2\ell}\cos\theta d\theta\wedge d\phi\right)\right]\,.
\end{split}
\]

Next we will prove that the above special Killing forms agree with the
special Killing forms $\Xi$ and $\Upsilon$ recently obtained  in
\cite{S-V-V,Vis}.
Indeed, if we denote
\[
p(y)=\frac{2y^3-3y^2+a}{3(1-y)},
\]
then we can easily remark that
\begin{equation}\label{sup1}
a(y)p(y)=-\frac{1}{2\ell}
\end{equation}
and
\begin{equation}\label{sup2}
\sqrt{y^3-\frac{3}{2}y^2+\frac{a}{2}}a(y)=-\frac{3}{2\ell}\sqrt{\frac{1-y}{
6p\left(y\right)}}.
\end{equation}

Using now (\ref{ag}), (\ref{sup1}) and (\ref{sup2}), we obtain that the real
special
Killing forms $\Re \Psi$ and $\Im \Psi$ can be rewritten as:
\begin{equation}\label{RePsi}
\begin{split}
&\Re \Psi  =
-\frac{3}{2\ell}\sqrt{\frac{1-y}{6p\left(y\right)}}\left[
\cos\psi\left(d\theta\wedge dy
-p(y)\sin\theta d\psi\wedge d\phi+6p(y)\sin\theta d\phi\wedge
d\alpha\right)\right.\\
& \left. ~~~~-\sin\psi\left(-p(y) d\theta\wedge d\psi-6p(y)d\theta\wedge
d\alpha-\sin\theta dy\wedge d\phi+
p(y)\cos\theta d\theta\wedge d\phi\right)\right]\,,
\end{split}
\end{equation}
\begin{equation}\label{ImPsi}
\begin{split}
&\Im \Psi  = -\frac{3}{2\ell}\sqrt{\frac{1-y}{6p\left(y\right)}}\left[\sin\psi
\left(d\theta\wedge dy
-p(y)\sin\theta d\psi\wedge d\phi+6p(y)\sin\theta d\phi\wedge
d\alpha\right)\right.\\
& \left. ~~~~+\cos\psi\left(-p(y) d\theta\wedge d\psi-6p(y)d\theta\wedge
d\alpha-\sin\theta dy\wedge d\phi+
p(y)\cos\theta d\theta\wedge d\phi\right)\right]\,.
\end{split}
\end{equation}

Therefore, because $\Re \Psi$ and $\Im \Psi$ coincide with $\Xi$ and $\Upsilon$
modulo a
multiplicative constant, we conclude that indeed the special Killing forms
obtained in
this article agree with the results previously obtained in \cite{S-V-V,Vis}
with different approaches.

\section{Conserved quantities and complete integrability of geodesic motion
in $Y^{p,q}$ spaces}

On a manifold with coordinates $x^\mu$ and metric $g_{\mu\nu}$ the geodesics
can be
defined as the trajectories of test-particles with proper-time Hamiltonian
\begin{equation}\label{Ham}
H = \frac12 g^{\mu\nu} P_\mu P_\nu \,.
\end{equation}
Here $P_{\mu}$ are canonical momenta conjugate to the coordinates $x^\mu$,
$P_\mu = g_{\mu\nu}\dot{x}^\nu$ with overdot denoting proper time derivative.

The system of a free particle admits conserved quantities \eqref{SKcons}
which commute with the Hamiltonian \eqref{Ham} in the sense of Poisson brackets:
\begin{equation}\label{PB}
\{K,H\} = 0\,.
\end{equation}

Let us recall that in classical mechanics a Hamiltonian system with Hamiltonian
$H$ \eqref{Ham} and integrals of motion $K_j$ is called {\it completely
integrable}
(or Liouville integrable) if it allows $n$ integrals of motion $H, K_1,
\dots, K_{n-1}$ which are well-defined functions on the phase space, in
involution
\begin{equation}
\{H, K_j\} = 0\,,~~ \{K_j, K_k\} = 0\,,\ ~~j,k = 1,\cdots , n-1\,,
\end{equation}
and functionally independent. A system is {\it superintegrable} if it is
completely integrable and allows further functionally independent integrals of
motion.

For $Y^{p,q}$ spaces the conjugate momenta to the coordinates
$(\theta,\phi, y, \alpha, \psi)$ are \cite{BK}:
\begin{equation}\label{momenta}
\begin{split}
&P_{\theta} =
\frac{1-y}{6} \dot{\theta}\,,\\
&P_{\phi} + \cos\theta P_{\psi} = \frac{1-y}{6} \sin^2\theta \dot{\phi}\,,\\
&P_y = \frac{1}{6 p(y)} \dot{y}\,,\\
&P_{\alpha}=w(y) \left(\dot{\alpha} + f(y)  \left(\dot{\psi} - \cos\theta
\dot{\phi}\right)\right) \,,\\
&P_{\psi} = w(y) f(y) \dot{\alpha} +
\left[ \frac{q(y)}{9} + w(y) f^2(y)\right]\left(\dot{\psi} - \cos\theta
\dot{\phi}\right)\,.
\end{split}
\end{equation}

From the isometry $SU(2) \times U(1) \times U(1)$ of the metric \eqref{Ypq}
we have that the momenta $P_\phi, P_\psi$ and $P_\alpha$ are conserved.
$P_\phi$ is the third component of the $SU(2)$ angular momentum and
$P_\psi, P_\alpha$ are associated to the $U(1)$ factors. In addition,
the total $SU(2)$ angular momentum
\begin{equation}
\vec{J}^{~2} =P_{\theta}^2 + \frac{1}{\sin^2\theta} \left(P_{\phi}+
\cos\theta P_{\psi}\right)^2 + P_{\psi}^2  \,
\end{equation}
is also conserved \cite{BK,RS}.

The next conserved quantities, quadratic in momenta, will be expressed
in terms of St\"{a}ckel-Killing tensors as in \eqref{SKcons}. The
St\"{a}ckel-Killing tensors of rank two on $Y^{p,q}$ will be constructed
from Killing-Yano tensors according to \eqref{KYY}. For this purpose we
shall use the Killing-Yano tensor $\Psi_1$ from \eqref{sKPsi}
and the additional parallel forms of degree $2$, associated with the real
and imaginary parts of the holomorphic $(3,0)$ volume form
$\Omega$ of the cone $C(Y^{p,q})$.

The first St\"{a}ckel-Killing tensor $K^{(1)}_{\mu\nu}$ is constructed
according to
\eqref{KYY} using the real part of the Killing form $\Psi$ \eqref{RePsi}:
\begin{equation}\label{K1RR}
K^{(1)}_{\mu\nu} =(\Re\Psi)_{\mu\lambda}
(\Re\Psi)_{\phantom{\lambda}\nu}^{\lambda}\,.
\end{equation}
The corresponding conserved quantity \eqref{SKcons}\footnote{In \cite{RS} the
expression of this conserved quantity has
some misprints. Consequently, the evaluation of the number of functionally
independent set of integrals of motion is affected and the system is not
superintegrable.}, modulo a multiplicative constant,
is \cite{BV}
\begin{equation}\label{K1}
\begin{split}
K^{(1)} =& 6(1-y)\dot{\theta}\dot{\theta}  + \frac{3 +a -6y +2y^3 +
(-3 +a +6y - 6y^2 + 2 y^3)\cos 2\theta}{1-y}\dot{\phi}\dot{\phi}\\
&-24\frac{(a + (-3+2y)y^2)\cos\theta}{1-y}\dot{\phi}\dot{\alpha}
- 4\frac{(a + (-3+2y)y^2)\cos\theta}{1-y}\dot{\phi}\dot{\psi}\\
&+ 18\frac{1-y}{a + (-3 +2y)y^2}\dot{y}\dot{y}
+72\frac{a + (-3+2y)y^2}{1-y}\dot{\alpha}\dot{\alpha}\\
&+24\frac{a + (-3+2y)y^2}{1-y}\dot{\alpha}\dot{\psi}
+2\frac{a + (-3+2y)y^2}{1-y}\dot{\psi}\dot{\psi}\,.
\end{split}
\end{equation}

The next St\"{a}ckel-Killing tensor  will be constructed from the imaginary
part of
$\Psi$ \eqref{ImPsi}:
\begin{equation}\label{K2II}
K^{(2)}_{\mu\nu} =(\Im\Psi)_{\mu\lambda}
(\Im\Psi)_{\phantom{\lambda}\nu}^{\lambda}\,,
\end{equation}
and we find that this tensor produces the same conserved quantity $K^{(1)}$
\eqref{K1}.

The mixed combination of $\Re \Psi$ and $\Im \Psi$ produces the
St\"{a}ckel-Killing
tensor
\begin{equation}\label{K3RI}
K^{(3)}_{\mu\nu} =(\Re\Psi)_{\mu\lambda}
(\Im\Psi)_{\phantom{\lambda}\nu}^{\lambda}
+ (\Im\Psi)_{\mu\lambda} (\Re\Psi)_{\phantom{\lambda}\nu}^{\lambda}\,,
\end{equation}
but it proves that all components of this tensor vanish.

Finally we construct the St\"{a}ckel-Killing tensor from the Killing form
$\Psi_1$:
\begin{equation}\label{K4PP}
K^{(4)}_{\mu\nu} =(\Psi_1)_{\mu\lambda\sigma}
(\Psi_1)_{\phantom{\lambda\sigma}\nu}^{\lambda\sigma}\,.
\end{equation}
From \eqref{sKPsi} using the $1$-form $\eta$ \eqref{eta} we get
\begin{equation}\label{Psi1}
\begin{split}
\Psi_1 & = (1-  y)^2\sin\theta \, d\theta \wedge d\phi \wedge
d\psi - 6 dy \wedge d \alpha \wedge d \psi \\
& \quad + 6 \cos\theta \,
d\phi\wedge dy \wedge d\alpha - 6 (1-  y) y
\sin\theta \, d\theta \wedge d\phi \wedge d\alpha\,.
\end{split}
\end{equation}
and the corresponding conserved quantity, modulo a multiplicative constant, is
\cite{BV}
\begin{equation}\label{K4}
\begin{split}
& K^{(4)}=6(1-y)\dot{\theta}\dot{\theta}
- 24\frac{(a + (-4 +5y -2y^2)y)\cos\theta}{1-y}\dot{\phi}\dot{\alpha}\\
&+\frac{7\! +\!a \!-\!18y\! + \!12 y^2\! -\!2y^3 \!+ \!
(1\! +\!a\!-\!6y\!+\!6y^2\!-\!2 y^3)\cos 2\theta}{1-y}\dot{\phi}\dot{\phi}\\
&- 4\frac{(a -(2-y)^2 (-1+2y))\cos\theta}{1-y}\dot{\phi}\dot{\psi}\\
&+ 18\frac{1-y}{a + (-3 +2y)y^2}\dot{y}\dot{y}
+72\frac{a + (1-2y)y^2}{1-y}\dot{\alpha}\dot{\alpha}\\
& +24\frac{a + (-4+5y-2y^2)y}{1-y}\dot{\alpha}\dot{\psi}
+2\frac{a -(2-y)^2 (-1+2y)}{1-y}\dot{\psi}\dot{\psi}\,.
\end{split}
\end{equation}

Having in mind that $K^{(1)}=K^{(2)}$ and $K^{(3)}$ vanishes, we shall verify
if the set
$H, P_{\phi},P_{\psi}, P_{\alpha}, \vec{J}^{~2}, K^{(1)}, K^{(4)}$ constitutes a
functionally independent set of constants of motion for the geodesics of
$Y^{p,q}$ constructing the Jacobian:
\begin{equation}
\mathcal{J} = \frac{\partial(H, P_{\phi},P_{\psi}, P_{\alpha},
\vec{J}^{~2}, K^{(1)}, K^{(4)})}
{\partial(\theta,\phi, y, \alpha,\psi,
\dot{\theta}, \dot{\phi}, \dot{y}, \dot{\alpha}, \dot{\psi})}\,.
\end{equation}
The rank of this Jacobian is $5$, exactly the number of the degrees of freedom,
which means that the system is completely integrable.
In spite of the presence of the St\"{a}ckel-Killing tensors $K^{(1)}$ and
$K^{(4)}$,
the system is not superintegrable, $K^{(1)}$ and $K^{(4)}$ being a combination
of the
first integrals $H, P_{\phi},P_{\psi}, P_{\alpha}, \vec{J}^{~2}$.
It is interesting to note that the toric Sasaki-Einstein spaces $Y^{p,q}$
spaces possess several Killing-Yano tensors, but these Killing forms do not
generate
new St\"ackel-Killing tensors, i.e. genuine conserved quantities.

\section{Conclusions}

In this article  we investigate the complex structure of the conifold
$C(Y^{p,q})$
basically making use of the interplay between symplectic and complex approaches
of
the K\"ahler toric manifolds. The description of the Calabi-Yau manifold
$C(Y^{p,q})$
with toric data allows us to write explicitly the complex coordinates and
special
Killing forms. Using the complete set of Killing vectors and St\"ackel-Killing
tensors
on $Y^{p,q}$ we construct the corresponding conserved quantities and proved the
complete
integrability of geodesic motion.

The present investigation is important in the context of the AdS/CGT
correspondence.
By focusing on the geodesics on the Sasaki-Einstein spaces, the paper refers to
geometries produced by $D$-branes on non-flat bases going towards the general
goal
of classifying all supersymmetric geometries with integrable geodesics.
It is quite remarkable the fact that while the point-like strings (geodesic)
equations are integrable in some backgrounds, the corresponding extended
classical
string motion is not integrable in general \cite{BZ1,BZ2,ST,CL1,CL2}.

\section*{Acknowledgments}

MV is grateful to L. A. Pando Zayas, O. Lunin, E. M. Babalic for useful
comments.
The work of VS was supported by the UE grant FP7-PEOPLE-2012-IRSES-316338.
MV was supported by CNCS-UEFISCDI, project number PN-II-ID-PCE-2011-3-0137.
The work of GEV  was supported by CNCS-UEFISCDI, project number
PN-II-ID-PCE-2011-3-0118.

\end{document}